\documentclass[pre,twocolumn, showpacs, preprintnumbers, amsmath,
amssymb]{revtex4-1}
\usepackage{graphicx}
\usepackage{hyperref}
\usepackage{amssymb}
\usepackage{amsmath}
\usepackage{dcolumn}% Align table columns on decimal point
\usepackage{bm}% bold math

\begin{document}

\title{Boundary mobility controls glassiness of confined colloidal liquids}

\author{Gary L. Hunter$^{1,2}$}
\email{GLHunter@gmail.com}
\author{Kazem V. Edmond$^{1,2}$}
\author{Eric R. Weeks$^{1}$}
\address{$^1$Department of Physics, Emory University, Atlanta, GA 30322, USA}
\address{$^{2}$Current address:  Center for Soft Matter Research, Department of Physics, New York University, New York, NY 10003, USA}

\date{\today}

\begin{abstract}
We use colloidal suspensions encapsulated in emulsion droplets to model confined glass-forming liquids with tunable boundary mobility. We show dynamics in these idealized systems are governed by physical interactions with the boundary. Gradients in dynamics are present for more mobile boundaries, whereas for less mobile boundaries gradients are almost entirely suppressed. Motions in a system are not isotropic, but have a strong directional dependence with respect to the boundary. These findings bring into question the ability of conventional quantities to adequately describe confined glasses. 
\end{abstract}
\pacs{64.70.pv, 61.43.Fs, 82.70.Dd}
%%% ----------------------------------------------------------------------
\maketitle

%%%%%%%%%%%%%%%%%%%

When cooled quickly, some liquids avoid crystallization and vitrify, becoming mechanically solid-like glasses~\cite{hunter12review}.  The glass transition temperature $T_{\rm g}$ marks the point where molecular motions all but cease.  For polymers and molecular liquids, one typically quantifies a material's glassiness in terms of relaxation times $\tau$, which behave roughly as the inverse of mobility and increase dramatically as the material is cooled toward $T_{\rm g}$.  For bulk glasses, these statements are sufficiently general and apply universally.   However, for small systems $T_{\rm g}$ and $\tau(T)$ exhibit a dependence on system size and surprisingly these quantities may either increase or decrease relative to their bulk values~\cite{lowen95,richert94,richert96,he07,richert11,pye2010,mattson2000,bohme02,mckenna05,paeng2012,roth07,torkelson05,ellison03,kob2002interface,kob00cylinder,kob04,nugent2007prl,eral09,zhu11,nabiha2013}.  These size-dependent variations in material properties are often collectively referred to as ``confinement effects'' and the understanding of such effects is essential to the development of functional nanoscale materials.

Research from the polymer and molecular liquid communities shows that the manner in which confinement affects mobility is dependent on the material in which the glass is confined~\cite{richert94,richert96,mattson2000,bohme02,kob2002interface,kob00cylinder,he07,richert11,kob04,mckenna05,roth07,pye2010,paeng2012,torkelson05,ellison03}.     In particular, the interaction (or lack of interaction) between the sample and the boundary is important.  Solid boundaries that chemically bond to the confined material suppress molecular mobility within the sample, whereas mobile or chemically repulsive boundaries enhance mobility~\cite{bohme02,mattson2000,pye2010,paeng2012,richert96,mckenna05,he07,richert11,roth07}.  Experimental evidence supports models in which these types of boundary effects propagate into a sample and give rise to gradients in dynamics~\cite{bohme02,mattson2000,pye2010,paeng2012,richert96,mckenna05,he07,richert11,roth07}. The idea is that confining a sample to a smaller space results in a large proportion of the sample being close to the boundary and therefore, as the system is made smaller, $T_{\rm g}$ and $\tau(T)$ are more statistically influenced by material near the boundary.  Therefore, depending on the nature of the interactions, these quantities may increase or decrease. Despite the evidence for dynamical gradients, it remains unclear how boundary mobility affects motions in different directions, i.e. in directions tangential or perpendicular to the boundary~\cite{frank96,tseng00,roth05}.

Colloidal suspensions have served as a valuable model of supercooled liquids and glasses and have elucidated much of the fundamental physics underlying the glass transition in molecular systems (see Ref.~\cite{hunter12review} and references therein). In hard-sphere colloids, phase behavior is controlled by volume fraction $\phi$ rather than temperature (qualitatively, $\phi~\sim~1/T$).  When confined within rigid, immobile boundaries, otherwise liquid-like colloidal suspensions transition to glassy dynamics at lower $\phi$ than in bulk samples where $\phi_{\rm g} \approx 58\%$, comparable to an increase in $T_{\rm g}$ for molecular glasses~\cite{nugent2007prl,eral09,zhu11,edmond10b}.  These reports also show that the length scales at which samples become glassy increase with increasing $\phi$.  Dynamics in confined colloids also depend on wall roughness~\cite{edmond10b,eral09}, a feature found in molecular dynamics simulations of Lennard-Jones liquids~\cite{kob04,kob00cylinder}.  To date however, no experiments with confined colloidal glasses have discussed the effect of directly modifying the mobility of the confining boundary. Hence it is unknown if the strong manner in which confined molecular liquids respond to boundary conditions is universal and also applies to colloids, or if the response depends on specific details of the system.

%THE SEE REF REFERENCE BELOW IS HUNTER12REVIEW 
%FIGURE 1
\begin{figure}
\includegraphics[width=0.43\textwidth]{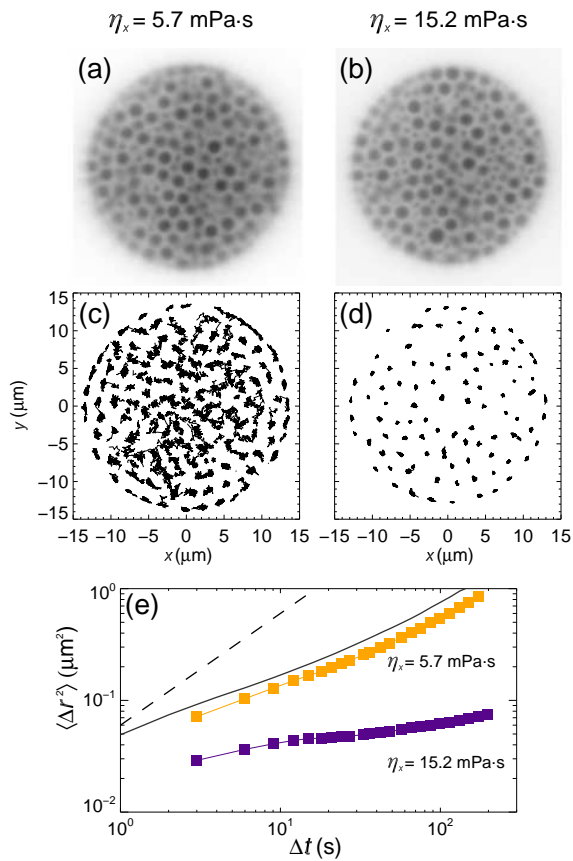}
 \caption{(Color online)  Colloidal suspensions encapsulated within emulsion droplets. (a,b), raw confocal images of confined colloids at $\phi = 46\%$ with (a) $\eta_x$ = 5.7 mPa$\cdot$s and $R$ = 14.4 $\mu$m and (b) $\eta_x$ = 15.2 mPa$\cdot$s and $R$ = 14.1 $\mu$m.  Movies of these data are included in the Supplemental Movie S1.  Note that these data are from separate experiments and that droplets are not physically near one another.  (c,d), Two-dimensional trajectories of the larger species of particles from the data immediately above.  Motions are shown over a period of 600 s.  Voids apparent in (c,d) are due to untracked particles slightly out of the focal plane of the microscope. Bulk translational motions are subtracted with standard particle tracking routines\cite{crocker1996jcis}, and bulk rotational motions are subtracted with a modified version of the procedures described in\cite{huntermethod11}.  (e), Mean-square displacements for the large particles in the droplets shown.  Light symbols (orange in color) correspond to data in (a,c).  Dark symbols (purple in color) correspond to data in (b,d).  The solid line is the MSD for an unconfined suspension at $\phi = 46\%$ and the dashed line has a slope of unity for comparison to Brownian diffusion.}
\label{fig:trajcomp}
\end{figure}

To address these issues directly, we use bidisperse colloidal suspensions confined in emulsion droplets as model glass-formers with tunable boundary conditions.  We observe these systems with fast confocal microscopy and use particle tracking methods determine the motions of many particles~\cite{crocker1996jcis,huntermethod11}.  We limit our attention to systems at volume fractions $\phi =$ 33 $\pm$ 1.5\% or $\phi =$ 46 $\pm$ 1.5\% (hereafter referred to as $33\%$ and $46\%$, respectively; details of uncertainties in $\phi$ are given in Ref.~\cite{dropletprlsuppmat}).  

The colloids are fluorescent poly(methyl methacrylate) (PMMA) spheres~\cite{andy2010} with small and large radii $a_{\rm S}$ = 0.532 $\mu$m and $a_{\rm L}$ = 1.08 $\mu$m.  The spheres are dispersed in a density- and index-matched solvent of cyclohexylbromide and decalin and are stabilized against aggregation by a $\approx$~15 nm layer of poly(12-hydroxy-stearic acid) (PHSA). Electrostatic repulsion between the particles is screened by saturating the solvent with tetrabutylammonium bromide salt~\cite{yethiraj03}. 

Colloid-filled droplets, such as shown in Fig.~\ref{fig:trajcomp}(a,b), are created by depositing small amounts of suspension onto mixtures of glycerol and water and shaking gently by hand.  A small amount of SDS surfactant (3mM) is present in the glycerol/water solution prior to shaking to stabilize the droplets against coalescence and prevent wetting of particles at the fluid-fluid interface (we do not observe the Pickering effect under these conditions).  The droplets are injected into glass chambers, allowed to sediment, and imaged via 2D or 3D confocal microscopy. The smaller particles move too quickly to be tracked reliably, so only trajectories for the large particles are computed~\cite{crocker1996jcis,huntermethod11}.  However, the numbers of small $n_{\rm S}$ and large $n_{\rm L}$ particles are accurately identified in each droplet.   Additionally, we find no evidence of particle size segregation over the course of several weeks.  Droplet radii $R$ are determined by measuring the well-defined radial coordinate of the outermost layer of large particles and adding $a_{\rm L}$.  Hence, volume fraction is given by $\phi = (n_{\rm S} a_{\rm S}^3 + n_{\rm L} a_{\rm L}^3) / R^3$~\cite{poon12}.   The number ratio $n_{\rm S}/n_{\rm L}$ varies slightly between droplets but is approximately $n_{\rm S}/n_{\rm L} \approx $ 0.9 $\pm$ 0.1 for $\phi = 33\%$ and 1.1 $\pm$ 0.2 for $\phi = 46\%$.  Precise values are given in Ref.~\cite{dropletprlsuppmat}. Only droplets with $R \le 18~\mu$m could be successfully observed due to the limited field of view of the confocal microscope and the index of refraction mismatch between the internal and external phases.  Droplet sizes are well-below the capillary length for our solvents ($\l_{\rm c} \approx 6$~mm $\gg R$), so droplets do not significantly deform under gravity. Thermal fluctuations in the droplet surfaces are $\approx 1$ nm, hence,  these droplets function as smooth, non-deformable spherical confining cells.  
%removed citation 

To vary mobility at the confining boundary, we change the viscosity of the external aqueous phase.  When a solitary particle of radius $a$ is suspended in an unbounded Newtonian fluid, it experiences a constant drag coefficient, $6\pi \eta a$.  Near a flat, mobile fluid-fluid interface however, the drag coefficient is a nontrivial function of distance from the interface and the viscosities of both the suspending and external fluids\cite{Yang1984Particle,Lee1979Motion}.  The qualitative behavior in such a situation is intuitive: increasing (decreasing) the viscosity of the external fluid increases (decreases) the drag coefficient, provided that the particle is within a distance of $\approx 10a$ of the interface.  Thus, by varying the viscosity of the external continuous phase, $\eta_x$, one modifies the viscous hydrodynamic coupling across the fluid-fluid interface and directly affects particle diffusivity. While the prior theory\cite{Yang1984Particle,Lee1979Motion} was developed for an isolated particle near a flat fluid-fluid interface, one expects the qualitative picture to remain the same for denser suspensions and curved interfaces.  

This is indeed what we observe, as shown in Fig~\ref{fig:trajcomp}(c,d) comparing the motions of particles in droplets of similar size and $\phi$, but where $\eta_x$ differs by a factor of 2.7.   Motion near the droplet interface is faster when $\eta_x$ is smaller (Fig.~\ref{fig:trajcomp}c); more surprising however is the dramatic difference in particle motion far from the boundary, where any direct hydrodynamic influence of the wall is screened~\cite{bradywall}. Particle mean-square displacements (MSDs) given in Fig.~\ref{fig:trajcomp}(e) show the striking difference in the magnitude of particle motions in these two droplets and demonstrate that the ease with which particles move at the boundary influences dynamics in the whole system.
%\cite{bradywall
% 

\begin{figure}
%\centering
\includegraphics[width=0.43\textwidth]{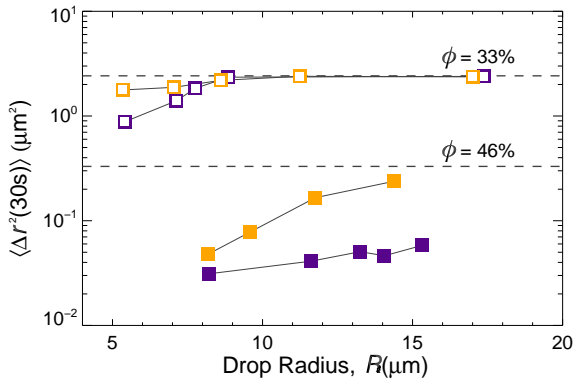}
 \caption{(Color online) MSDs at $\Delta t$ = 30 s of particles in droplets of different $\phi$ and different~$\eta_x$ as a function of drop radius.  Open symbols correspond to $\phi = 33\%$ and solid symbols are for $\phi = 46\%$.  Light symbols (orange in color) are for $\eta_x = 5.7$ mPa$\cdot$s and dark symbols (purple in color) are for $\eta_x = 15.2$ mPa$\cdot$s.  Dashed horizontal lines are  $\langle \Delta r^2(\rm{30 s}) \rangle$ in unconfined samples at the same $\phi$.  Data with $\phi = 46\%$ are from two-dimensional confocal images at the equatorial plane of the droplets whereas data with $\phi = 33\%$ are from three-dimensional confocal images of the entire droplets.  For meaningful comparison, we scale data for $\phi = 33\%$ such that $\langle \Delta r^2 \rangle = \frac{2}{3} \langle \Delta x^2 +\Delta y^2 + \Delta z^2 \rangle $ whereas for data at higher $\phi$, $\langle \Delta r^2 \rangle = \langle \Delta x^2 +  \Delta y^2 \rangle$ . }\label{fig:msds}
\end{figure}

The effect of $\eta_x$, as well as $\phi$ and $R$, are more clearly represented by examining the MSDs at a single lag time $\Delta t$.  MSDs as a function of $\Delta t$ are given for all data sets in Ref.~\cite{dropletprlsuppmat}.  As shown there, MSDs vary quantitatively with $\Delta t$, but the behavior as a function of $R$ or $\phi$ remains qualitatively the same for any $\Delta t$.  Shown in Fig.~\ref{fig:msds} are particle MSDs at $\Delta t$ = 30 s as a function of droplet radius. For $\phi = 33\%$ and droplet radii $R \gtrsim 9~\mu$m, dynamics are indistinguishable from those in an unconfined, bulk sample.  Therefore, no obvious effect of confinement or $\eta_x$ is present at these length scales.  Below this size, we observe decreased particle mobility and the onset of confinement effects.

%FIGURE 2

With decreasing droplet size we find further reduction in particle mobility, however most significantly, the rate at which dynamics decrease with $R$ differs between the two $\eta_x$, with motions decreasing more strongly for larger $\eta_x$ (dark symbols in Fig.~\ref{fig:msds}).  For samples at $\phi = 46\%$, mobilities are less than those in bulk for all droplets observed, therefore the onset of a confinement effect occurs at droplet sizes $R \gtrsim 15~\mu$m.  Hence, as $\phi$ increases so do length scales associated with the onset of confinement effects.  This observation relates to growing length scales near the glass transition~\cite{hunter12review} and is consistent with previous research on colloidal suspensions confined in rigid chambers with immobile boundaries\cite{nugent2007prl,edmond10b, nabiha2013}.    We also find that for $\phi > 50\%$ (data not shown), particle motions for our range of observable $R$ are very small, and are on the scale of our uncertainty in tracking.  As with data at lower $\phi$, the rates at which dynamics slow with droplet size depend on $\eta_x$, indicating that particles within the droplet are indeed affected by properties outside or at the fluid-fluid interface.   In general after the onset of confinement effects, particles in larger droplets move faster than those in smaller droplets, and a higher viscosity external phase results in lower particle mobility.  The change in mobility with $\eta_x$ is reminiscent of observations in confined polymers and small molecule glasses, where $T_{\rm g}$ and $\tau$ are strongly dependent on properties of the confining interface~\cite{roth07,pye2010,paeng2012,torkelson05,ellison03,richert94,richert96,he07,richert11,kob04,kob2002interface,kob00cylinder}.

Changes in $T_{\rm g}$ and $\tau(T)$ (increases or decreases) relative to the bulk have been ascribed to mobility gradients originating from interactions at the boundary\cite{bohme02,mattson2000,pye2010,paeng2012,richert96,mckenna05,he07}. Visualizing the confined particles with confocal microscopy allows such gradients to be observed directly.  Individual particle MSDs are resolved into radial and angular components, and the mobility computed as $\Delta r = \sqrt{\langle \Delta r_r^2 + \Delta r_{\theta}^2 \rangle}$ as a function of a particle's distance $s$ from the droplet interface (see Ref.~\cite{dropletprlsuppmat} for further details of the calculation).  We limit our discussion of mobility gradients to experiments where $\phi = 46\%$, but results from the lower $\phi$ case are similar and are given in Ref.~\cite{dropletprlsuppmat}.  

%FIGURE 3
%can use \begin{figure*} to force full width fig.
\begin{figure}
\centering
\includegraphics[width=0.46\textwidth]{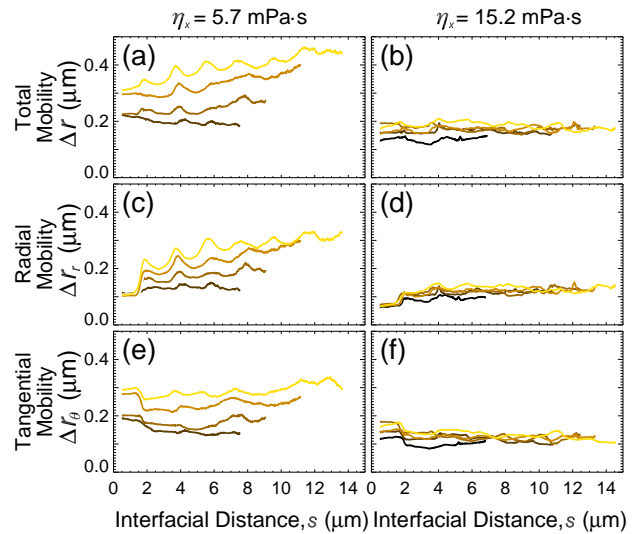}
 \caption{(Color online) Mobility as a function of interfacial distance.  (a,b) Particle mobility $\Delta r = \sqrt{\Delta r_r^2 + \Delta r_{\theta}^2}$ as a function of distance $s$ from the fluid-fluid interface for systems with $\phi = 46\%$ and at $\Delta t = 30$ s. Data are from two-dimensional confocal images at equatorial plane of the droplets and so include only one angular direction tangential to the interface.  Data in the left panels (a,c,e) have $\eta_x = 5.7$~mPa$\cdot$s and right panels (b,d,f) have $\eta_x = 15.2$~mPa$\cdot$s. Droplets decrease in size from light to dark (orange to black in color). Where the data terminates at larger distances is approximately the droplet radius.  See Tabs. II and III in Ref.~\cite{dropletprlsuppmat} for precise radii. (c,d), Radial  components $\Delta r_r$ and (e,f), tangential components $\Delta r_{\theta}$ of particle mobility.}  \label{fig:gradshigh}
\end{figure}
%\end{widetext}

% THE SEE REF REFERENCES BELOW ARE nugent2007prl,edmond10b,eral09
Shown in Fig.~\ref{fig:gradshigh}(a,b) are particle mobilities $\Delta r$ as a function of interfacial distance $s$ for droplets of different sizes and different $\eta_x$.   Dynamical gradients are apparent in droplets with the lower viscosity outer phase, and the slope of the gradient decreases with decreasing droplet size, perhaps even becoming negative for the smallest droplet studied.  For the larger droplets, particles are generally slower at the boundary than in the center of the droplet.  With decreasing droplet size, mobility curves shift to lower values, as would be expected from Fig.~\ref{fig:msds}. In the case of the higher $\eta_x$, there are no obvious indications of a mobility gradient, but the decrease in mobility with decreasing droplet size is present.  Small oscillations in mobility are the result of density fluctuations due to particle layering and are a generic feature of confined particles\cite{nugent2007prl,mittal08,edmond10b,nabiha2013}.  Layering has been shown to have a significantly smaller effect on mobility than confinement (see Refs.~\cite{nugent2007prl,edmond10b,eral09} for detailed discussions of this effect).  The average structural properties of comparable droplets is independent of $\eta_x$ (see Ref.~\cite{dropletprlsuppmat}) and so does not explain differences in mobility.
%removed citation mittal2007jpcb2 in relation to layering

Decomposing mobility into radial and tangential components reveals further similarities between systems with different $\eta_x$.  In Fig.~\ref{fig:gradshigh}(c,d), radial motions near the interface are quite small for all droplets in a given $\eta_x$, but the mobilities of particles nearest the interface do exhibit a slight dependence on droplet size.  Approaching the center of the droplet, radial mobilities increase, though this observation is much more prominent in the lower $\eta_x$ case.   The increase of radial mobility far from the interface is similar to previous findings for the component of mobility perpendicular to a smooth or rough rigid boundary\cite{nugent2007prl, mittal08,edmond10b,nabiha2013,kob2002interface,kob00cylinder,kob04}, but variations of this type in response to boundary mobility have not been reported.

In contrast, the tangential mobilities  shown in Fig.~\ref{fig:gradshigh}(e,f) appear highest for the layer of particles immediately adjacent to the droplet interface ($s \leq 2a_{\rm L} \approx 2~\mu$m).  At $s\approx 2~\mu$m, the mobility falls in a stepwise fashion and remains essentially constant in the remainder of the droplet, with some larger fluctuations near the droplet center where statistics are poorer.  As with the radial component, a decrease in droplet size results in decreased average mobility and is a trend present for droplets in either external phase, but more pronounced when the external viscosity is lower.  Given the observations in Fig.~\ref{fig:gradshigh}(c-f) for systems with different $\eta_x$, the qualitative similarity combined with the stark quantitative difference points to the strong effect that boundary mobility can have on the dynamics of the entire system.

The functional form of the mobility gradient in our case is unclear. Several models whose dynamics vary continuously with interfacial distance have been substantiated~\cite{pye2010,he07,richert11,kob04,kob2002interface} but do not adequately describe our data. Prior models of confined glassy materials assume that molecular mobility is a function of $T$, distance from the boundary, and the boundary conditions~\cite{mattson2000,paeng2012,pye2010,kob2002interface,he07,richert11,kob04}.  Hence, more confined samples have a larger fraction of their material close to a boundary and the influence of the boundary dominates the sample-averaged dynamics.  Data in Fig.~\ref{fig:gradshigh} show that mobility in our samples depends on these factors (a dependence on $\phi$ rather than $T$), but strikingly, also depends on $R$.  For example, the data in Fig.~\ref{fig:gradshigh}(a) for $s<6~\mu$m vary appreciably with the overall droplet size $R$.  Thus, the slowed motion in more-confined droplets appears not just as the result of a stronger interface effect but is also partially due to a finite size effect, perhaps arising from the more pronounced curvature of the smaller droplets.  Indeed, confining geometry has been found to be an important parameter in confined polymers~\cite{steinhart2010}.   While many differences exist between our idealized systems and polymers or supercooled molecular liquids, e.g. variations in particle geometries and particle-particle interactions, the response of a confined material to boundary conditions appears somewhat universal.  Whether a general form incorporating appropriate variables can successfully describe the variety of confinement scenarios remains an open question.
%
%%FIGURE4
%\begin{figure}
%%\centering
%%\includegraphics[width=0.85\textwidth]{figures/sidebyside.eps}
%\includegraphics[width=0.47\textwidth]{figures/render1.eps}
% \caption{(Color online) Mobility of colloidal particles at $\phi = 46\%$ confined within a $\approx$ 11.8~$\mu$m radius emulsion droplet.  (a), Radial component of mobility and (b), tangential component at $\Delta t=$ 30 s.  The rendering is created using experimental particle positions and motions.  Radial mobility varies continuously as a function of interfacial distance, while tangential motions are most affected immediately adjacent to the interface.    Raw mobility data are shown in Fig.~\ref{fig:gradshigh}(c,e) as the outlined curves.}\label{gradrend}
%\end{figure}

Our data demonstrate that, as with polymers and molecular liquids, the properties of a medium external to a confined colloid can have a marked impact on dynamics, even relatively far from the boundary.  Thus, colloids continue to be a valuable model for glassy materials.  Because particle mobilities depend on $\eta_x$, it is clear that confinement effects in general are not merely the result of a system's finite size;  similar to polymers and small molecule glasses, the magnitude of motions occuring near the confining interface strongly influences motions in the remainder of the material.

An interesting implication of our observations relates to the interpretation of relaxation in confined glasses and how relaxation events and time scales in these scenarios can be appropriately quantified.  For isotropic motions in the bulk of a glassy material, these quantities appear well-defined.  However, if relaxation is anisotropic, occurring on different time scales in different directions, then conventional definitions of relaxation may need revision.  As found in prior experiments~\cite{nugent2007prl,edmond10b,nabiha2013} and simulations~\cite{kob04,mittal08}, motions parallel and perpendicular to the boundary depend strongly on the shape and roughness of the confining boundary.  Our results show that the fluidity of the boundary is an equally important parameter.  The directional dependence of particle mobilities implies that relaxation also occurs anisotropically.  Ignoring this anisotropy means that measurements of $\tau$ could be biased toward slower or faster motions (depending on the technique), obscuring or missing important physics entirely.   While extremely sensitive, techniques used in polymers (e.g. fluorescent response of dyes, fluorescence recovery after photobleaching) and molecular liquids (e.g dielectric relaxation, solvation dynamics) are not yet able to simultaneously distinguish directionally dependent relaxations.  In this respect, experiments with confined colloidal glasses may provide valuable insight for experiments with other confined glass-forming liquids.

%%%%%%%%%%%%%%%%%%%%%%%%%%%%%%%%%%%%%%%%
%\section*{Acknowledgements}
This work was supported by the National Science Foundation under Grant No. DMR-0804174. We thank G. C. Cianci for synthesizing particles and the M. J. Solomon group for donating the PHSA stabilizer.  We also thank S. H. Behrens, K. W. Desmond, L. Feng, A. Fernandez-Nieves, B. Laderman,  R. Richert, C. B. Roth,  and K. Warncke for valuable discussions.

%%%%%%%%%%%%%%%%%%%%%%%%%%%%%55

%\bibliography{droplets}

%Merlin.mbs v4.21 2009-07-09.
%

%%%%%%%%%%%%%%%%%%%%%%%%

%%END BIB
%Append in Bib
%

%%%%%%%%%%%%%%%%%%%%%%%%%%
%See Supplemental Material at [URL will be inserted by publisher] for precise droplet sizes, $\phi$, and.

\clearpage
%%%%%%%%%%%%%%%%%%%%%%%%%%%%%%%%%%%%%%%%%%%%

Boundary mobility controls glassiness of confined colloidal liquids

Authors:  Gary~L.~Hunter, Kazem~V.~Edmond, Eric~R.~Weeks

\section*{Supplemental Material}

\section*{Mean-square displacements}

\begin{figure}[h]
%\centering
\includegraphics[width=0.45\textwidth]{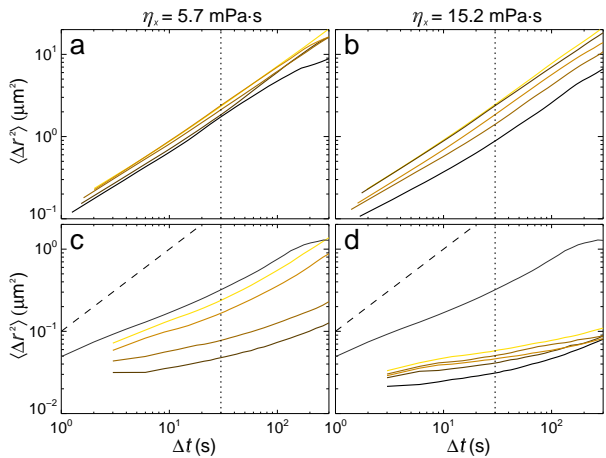}
 \caption{{\bf Particle mean square displacements in droplets of different sizes with different boundary conditions}. For left panels (a,c) $\eta_x$ = 5.7 mPa$\cdot$s and for right panels (b,d)  $\eta_x$ = 15.2 mPa$\cdot$s. For top panels (a,b) $\phi = 0.330 \pm 0.015$ and for bottom panels (c,d) $\phi = 0.460 \pm 0.015$. Note that data in (a,b) are from 3D data of the entire droplets and have been scaled as in Fig.~\ref{fig:msds}, i.e. $\langle \Delta r^2 \rangle = \frac{2}{3} \langle \Delta x^2 +\Delta y^2 + \Delta z^2 \rangle $ .The vertical dotted lines are placed at $\Delta t$ = 30 s for comparison with Fig.~\ref{fig:msds}.  For (a)-(d), droplet radii from top to bottom at $\Delta t$ = 30 s are: (a) 11.2, 17.0, 8.6, 7.0, 5.4 $\mu$m; (b) 8.9, 17.4, 7.8, 7.1, 5.4 $\mu$m; (c) 14.4, 11.8, 9.6, 8.2 $\mu$m;  (d) 15.3, 13.3, 14.0, 11.6, 8.2 $\mu$m.  See Tab.~\ref{tab:resultslow} for values of $R$.   For $\phi = 33\%$, unconfined samples have identical dynamics as the largest drops.  The dark solid line in (c,d) are unconfined data at $\phi = 46\%$.  Dashed lines indicate a slope of one.}  \label{fig:msds3345}
\end{figure}

The MSDs for all data sets discussed in the text are given in Fig.~\ref{fig:msds3345}.

\section*{Mobility gradients, $\phi = 33\%$}

As mentioned in the main text and shown in Fig.~\ref{fig:gradshigh}, apparent oscillations in mobility as a function of interfacial distance are the result of particles layering when confined\cite{nugent2007prl,mittal08,nabiha2013}. Hence, independent of any effect of $\eta_x$, a particle experiences different restrictions to motion depending on whether it is in a single layer or between layers.  Furthermore, particles occur less frequently between layers and statistics in these locations are poorer.  To better resolve the influence of the boundary and to smooth out fluctuations due to layering, we define a local average particle mobility  $\Delta r_e(\Delta t, s)$ as a function of distance $s$ to the boundary:

\begin{equation}
\Delta r_{e} (\Delta t, s) \equiv \langle | \Delta \vec{r} (\Delta t, s \pm \delta s) \cdot \hat{e} |\rangle.
\end{equation}
%
%\begin{equation}
%\Delta r_{e} (\Delta t, s) = \frac{\int\limits_{s-\delta s}^{s+\delta s}|\Delta \vec{r}(\Delta t, s) \cdot \hat{e}| \ n(s)s^{d-1}ds }{ \int\limits_{s-\delta s}^{s+\delta s} n(s)s^{d-1}ds},
%\end{equation}

\noindent The angle brackets $\langle \rangle$ indicate an average over all initial times $t$ for a fixed lag time $\Delta t$ and for all particles located within a range $s \pm \delta s$ from the boundary.  The unit vector $\hat{e}$ has direction $\hat{r}$ or $\hat{\theta}$ and is defined from a particle's position relative to the droplet center. The total mobility is simply the magnitude of a particle's vector displacement $|\Delta \vec{r}(s)|$, i.e. the square root of the MSD.  For consistency with Fig.~\ref{fig:msds}, we choose $\Delta t = 30$~s.  No qualitative difference is found using different $\Delta t$.  A value of $\delta s = 0.5~\mu$m is used for all data sets and is sufficiently small to reveal significant qualitative and quantitative trends, as shown in the text.  Thus, $\Delta r$ is a number-weighted average of the size of particle displacements over a radial bin with a width of $\approx$~1 large particle radius.

\begin{figure}[h]
%\centering
\includegraphics[width=0.45\textwidth]{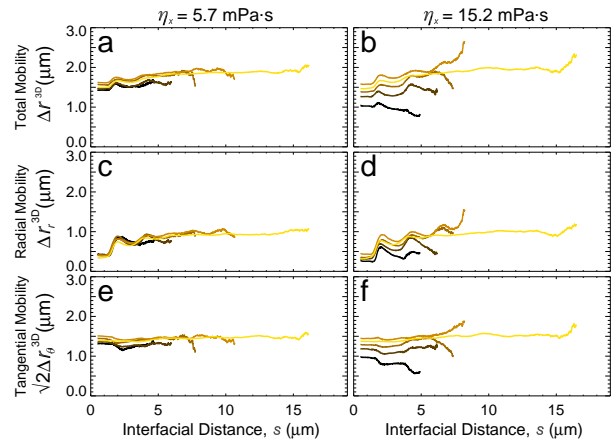}
 \caption{ (a,b) Particle mobility $\Delta r$ as a function of distance $s$ from the fluid-fluid interface for confined particles with $\phi = 33\%$. The 3D superscript refers to these data being from three dimensional confocal images of the entire droplets.  These results show the three dimensional mobility $\Delta r = \sqrt{\Delta r_r^2 + 2\Delta r_{\theta}^2}$, including both angular directions parallel to the interface.  Data in the left panels (a,c,e) have $\eta_x = 5.7$~mPa$\cdot$s and right panels (b,d,f) have $\eta_x = 15.2$~mPa$\cdot$s. Droplets have different radii and decrease in size from light to dark (orange to black in color). Where the data terminates at larger distances is approximately the droplet radius.  (c,d) Radial components $\Delta r_r$ and (e,f) tangential components $\sqrt{2}\Delta r_{\theta}$ of particle mobility. See Tabs. II and III for precise radii.} \label{fig:gradslow}
\end{figure}

\section*{Pair correlation functions, $\phi$ = 33\%}

\begin{figure}[h]
%\centering
\includegraphics[width=0.45\textwidth]{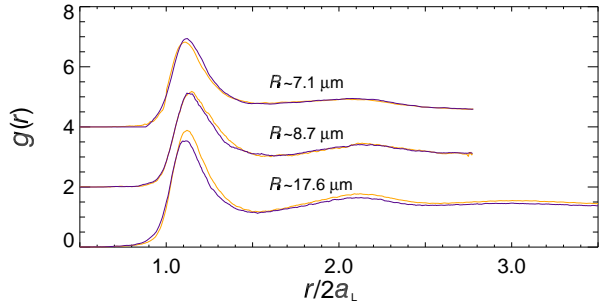}
 \caption{Comparison of pair correlation funtions of particles with $\phi = 33\%$ in droplets with similar size but different $\eta_x$: light curves (orange in color) are $\eta_x = 5.7$~mPa$\cdot$s; dark curves (purple in color) are $\eta_x = 15.2$~mPa$\cdot$s.    Here, distance $r$ on the horizontal axis is normalized by a large particle diameter, $2a_{\rm L}$ and truncated at $r/2a_{\rm L} \approx 2.8$ for very small droplets.  Curves have been vertically offset for clarity and labeled with the approximate droplet radii. See Tabs. II and III for precise radii.}  \label{fig:grslow}
\end{figure}

\begin{figure}
%\centering
\includegraphics[width=0.45\textwidth]{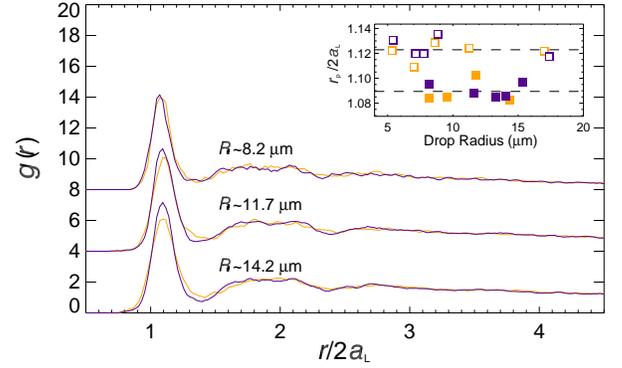}
 \caption{Comparison of pair correlation funtions of particles with $\phi = 46\%$ in droplets with similar size but different $\eta_x$: light curves (orange in color) are $\eta_x = 5.7$~mPa$\cdot$s; dark curves (purple in color) are $\eta_x = 15.2$~mPa$\cdot$s.  Here, distance $r$ on the horizontal axis is normalized by a large particle diameter, $2a_{\rm L}$ = 2.16~$\mu$m.  Note that curves have been vertically offset for clarity and labeled with the approximate droplet radii.  Inset: location of the first peak $r_{\rm p}$ in $g(r)$ for $\phi = 33\%$~(open symbols) and $\phi = 46\%$~(filled symbols). Symbol color has the same meaning as in the main figure.}  \label{fig:grs}
\end{figure}

As mentioned in the main text, we see no significant structural difference between droplets of similar size in the different external phases.  As shown in Figs.~\ref{fig:grslow} and \ref{fig:grs}, pair correlation functions $g(r)$ for comparable $\phi$ and $R$ are nearly identical.  Fluctuations in the locations $r_p$ of the peaks in $g(r)$, shown in the inset of Fig.~\ref{fig:grs}, are less than 2\% of a large particle diameter ($\lesssim 50$ nm), consistent with uncertainty in particle tracking\cite{crocker1996jcis,huntermethod11}.  At higher $\phi$, particles are packed more tightly, but $r_p$ exhibits no clear trend with droplet size or $\eta_x$.
%  \cite{fraden94} <-- peak of g(r) shifts with charge

%FIGURE5

\begin{table}
 %\centering 
\begin{tabular}{c c c} 
Composition & Density & \hspace{.5cm} $\eta_x$  \\
(w/w) & (g/mL) & \hspace{.5cm} (mPa$\cdot$s)\\
\cline{1-3}
 \\
50/50 & 1.124 & \hspace{.5cm} 5.7 \\
65/35 & 1.165 & \hspace{.5cm} 15.2\\
\hline
\end{tabular} 
\caption{Properties of External Glycerol/Water Phases}
\label{tab:props} 
\end{table}

%Results tables
\begin{table}
 %\centering 
\begin{tabular}{c c c c} 
$R$ & $n_{\rm S}/n_{\rm L}$ & $\phi$ & $\langle \Delta r^2({\rm 30 s}) \rangle$ \\
($\mu$m)  & & & ($\mu$m$^2$) \\
\cline{1-4}
 \\
      5.35    &         0.97  &   0.331 &     3.47\\
      7.03    &       0.98 &   0.336  &    3.39\\
      8.60    &      0.78   &  0.318  &   3.61\\
      11.24  &     1.00    &  0.327   &  3.83\\
      17.00  &   0.81   &  0.327   &   3.65\\
\hline
\\
8.18    &  1.21 &    0.459 &   0.048\\
      9.57 &     0.90 &    0.455 &   0.078\\
      11.76    &  1.31   &  0.457  &   0.166\\
      14.40 &     1.08 &    0.458 &    0.238\\
\hline
\end{tabular} 
\caption{Results of Confinement Experiments: $\eta_x$ = 5.7 mPa$\cdot$s. }
\label{tab:resultslow} 
\end{table}

%Results tables
\begin{table}[h]
 %\centering 
\begin{tabular}{c c c c} 
$R$ & $n_{\rm S}/n_{\rm L}$ & $\phi$ & $\langle \Delta r^2({\rm 30 s}) \rangle$ \\
($\mu$m)  & & & ($\mu$m$^2$) \\
\cline{1-4}
 \\
      5.40&       0.89&     0.327&      1.80\\
      7.13&     0.76&     0.338&      2.43\\
      7.76&       0.87&     0.345&      3.13\\
      8.86&      0.75&     0.315&      3.88\\
      17.37&      0.97&     0.334&      3.78\\
\hline
\\
         8.20  &    1.00 &    0.466 &   0.031\\
      11.63 &      1.11  &   0.470  &  0.041\\
      13.26 &     1.54 &    0.455 &   0.051\\
      14.06  &    1.04 &    0.459 &   0.046\\
      15.32   &   1.25 &    0.463 &   0.058\\
\hline
\end{tabular} 
\caption{Results of Confinement Experiments: $\eta_x$ = 15.2 mPa$\cdot$s. }
\label{tab:resultshigh} 
\end{table}

Relevant system parameters for all experiments presented here are summarized in Tabs.~\ref{tab:resultslow} and~\ref{tab:resultshigh}.  Droplet radii and MSDs are calculated from particle coordinates, hence uncertainties in these quantities are from particle tracking; $\delta R \approx 0.05~\mu$m and $\delta (\Delta r^2) \approx 0.003~\mu$m$^2$. We estimate that the number of small and large particles in a droplet are each determined to less than 1\% error, therefore the measurement uncertainty in the number ratio is $\delta (n_{\rm S}/n_{\rm L}) \approx 0.015$.  As stated in the main text, $\phi =  (n_{\rm S} a_{\rm S}^3 + n_{\rm L} a_{\rm L}^3) / R^3$.  Therefore, from standard error propagation, uncertainties in $\phi$ include contributions from uncertainty in the above parameters, as well as the polydispersity of the small and large particles ($\delta a/a \approx 0.04$). Including all contributions, $\delta \phi \approx 0.038$ and $\approx 0.051$ for lower and higher $\phi$, respectively.  Ignoring particle polydispersity, $\delta \phi \approx 0.010$ and $0.015$ at the two volume fractions.  Uncertainties for these parameters also apply to data with $\eta_x$ = 15.2 mPa$\cdot$s given in Tab.~\ref{tab:resultshigh}.  The variations quoted in the main text are meant to describe range of measured $\phi$ for droplets that are compared directly.

\end{document}